\documentclass[12pt]{iopart}

\usepackage{epsfig}
\usepackage{graphicx}
\usepackage{dcolumn}
\usepackage{bm}
\usepackage{times}
\renewcommand{\vec}[1]{{\bf #1}}

\begin{document}
\title{Tunneling of Dirac electrons through spatial regions of finite mass}
\author{J. Viana Gomes and N.~M.~R. Peres}
\address{Center of Physics  and  Department of Physics,
             University of Minho, P-4710-057, Braga, Portugal}

\begin{abstract}
We study the tunneling of chiral electrons in graphene through a
region where the electronic spectrum changes from the usual linear
dispersion to a hyperbolic dispersion, due to the presence of a gap.
It is shown that contrary to the tunneling through a potential
barrier, the transmission of electrons is, in this case, smaller
than one for normal incidence. This mechanism may be useful for 
designing  electronic devices made of graphene.
\end{abstract}

\submitto{\JPCM}

\section{Introduction}
A new and exciting field in condensed matter physics started when
graphene - a two-dimensional, one carbon-atom thick material - was
isolated for the first time.\cite{novo1,pnas} It was experimentally
shown that the charge carriers in graphene could be controlled by a
bottom gate setup-up; the charge carrier were shown to be either
holes or electrons depending on the sign of the bottom gate voltage.
In the transition from hole-based to electron-based transport the
conductivity shows a minimum (not zero) value, $\sigma_{\rm min}$.
Its experimental value is of the order of $\sigma_{\rm min}\simeq
4e^2/h$,\cite{novo1,pnas,novo3,kim3} but the actual value seems to
be some what sample dependent.\cite{Tan} This value for $\sigma_{\rm min}$
imposes therefore a limitation on the minimum value of the current a
field effect transistor made of graphene can transport. The
existence of a  conductivity  minimum in graphene is
 a consequence of the fact that the elementary excitations of graphene
are Dirac fermions, with a linear dispersion relation, instead of
usual electrons with parabolic-like dispersion, characteristic of
ordinary semi-conductors. Interestingly enough, the calculated value
of the conductivity of graphene at the neutrality point is off the
experimental value by the factor $1/\pi$.
\cite{Ando98,Ando02,Peres06}Although this value is the more common result
for the theoretical calculation of $\sigma_{\rm min}^{\rm theo }$,
 there are, however, several different
values available in the literature.\cite{Ziegler07} It is also
interesting that a clean graphene sample with metallic leads and
smooth edges has a value of $\sigma=\sigma_{\rm min}^{\rm theo}$ as
long as it width ($w$)  is much larger than its length ($L$), being
smaller
 than  $\sigma_{\rm min}^{\rm theo}$ in the opposite limit. \cite{Beenakker06}
Considering the case of metallic armchair edges, it found that
$\sigma>\sigma_{\rm min}^{\rm theo}$ for $w/L\ll 1$ and that
$\sigma\rightarrow\sigma_{\rm min}^{\rm theo}$ for $w/L\gg
1$.\cite{Beenakker06} This shows that disorder is not needed for
having $\sigma\simeq\sigma_{\rm min}^{\rm theo}$.

Another characteristic of Dirac electrons in graphene is their
ability to tunnel through a potential barrier with probability one.
\cite{NaturePhys,Bai07} This so called Klein tunneling of chiral
particles has long ago been proposed in the framework of quantum
electrodynamics,\cite{Klein29,Calogeracos,Zuber} but was never
observed experimentally. Graphene opens up a route to observe this
effect in a tabletop experiment, where the potential is created by
some electrostatic gate potential. The manifestation of Klein
tunneling is also present when electrons in graphene are forced to
transverse a $n-p$ junction, leading to a selective transmission of
those electrons approaching the junction
perpendicularly.\cite{cheianov06} Other unusual effects, such as the
focusing of an electric current by a single $p-n$ junction are also
characteristic of Dirac electrons in graphene. \cite{cheianov07}

As appealing as the Klein tunneling may sound from the point of view
of fundamental research, its presence in graphene is unwanted when
it comes to applications of graphene to nanoelectronics. This comes
about because the pinch-off the field effect transistor may be very
ineffective. The same may occur because of the minimum conductivity
of graphene at the neutrality point (as discussed above). One way to
overcome these difficulties is by generating a gap in the spectrum.
From the point of view of Dirac fermions this is equivalent to the
generation of a mass term. There are two known forms of generating
gaps in the spectrum of graphene. The first one is by patterning
graphene nanoribbons.\cite{Louie06,Duan07} The mechanism of
producing these gaps depends on the nature of the termination of
these nanoribbons. For armchair nanoribbons the gap comes from
quantum confinement of Dirac fermions induced by the finite nature
of the ribbons in the transverse direction. For zigzag nanoribbons
the gap stems from the formation of polarized spin edge-states
characteristic of these type of ribbons. The formation of these
polarized states is also possible in bilayer
graphene.\cite{Eduardo07} It is interesting to notice that Klein tunneling
can also be circunvented by using a graphene bilayer.\cite{NaturePhys}
The value of the induced gaps depend on
the width of the ribbons, but for large widths it is of the order of
0.1 {\ttfamily eV}.

Another possibility of generating gaps in the graphene spectrum is
to deposit graphene on top of hexagonal boron nitride (BN).
\cite{Giovannetti07} This material is a band gap insulator with a
boron to nitrogen distance of the order of 1.45 \AA, \cite{Zupan71}
(in graphene the carbon-carbon distance is 1.42 \AA) and a gap of
the order of $4$ {\ttfamily eV}. It was shown that in the most
stable configuration, where a carbon  is on top of a boron and the
other carbon in the unit cell is centered above a BN ring, the value
of the induced gap is of the order of 53 m{\ttfamily eV}. Depositing
graphene on a metal surface with a BN buffer layer leads  to $n-$doped
graphene with an energy gap of 0.5 {\ttfamily eV}.\cite{Lu07}

The two mechanisms described above can be used to produce
arrangements of graphene where in some spatial zones of the material
the Dirac electrons will have gaps in the spectrum. The first
possibility is to pattern graphene planes such that in several areas
of the graphene flake narrow nanoribbons may exist. Another
possibility it to combine wafers of silicon oxide and hexagonal
boron nitride, such that in the region where the BN is located the
local spectrum of graphene will present a finite gap. We shall
explore in this paper this latter possibility and the way it can
prevent Klein tunneling from occurring. The  two mechanisms just
mentioned can then be at the heart of future nanoelectronics built
of graphene. The second method is also related to junctions of
graphene with other kind of systems, being them
superconducting,\cite{Titov06,Moghaddam}
 normal-conductor/graphene/normal-conductor,
\cite{Robinson07} or multiterminal junctions.\cite{Jayasekera} Also
the study of electron transport in disordered graphene samples is of
interest,\cite{Wakabayashi} specially because the tunneling may be
assisted by impurities,\cite{Titov07} which is a manifestation of
Klein tunneling.

 For a review on the experimental aspects of graphene
physics see the work of Geim and Novoselov.\cite{Nov07} Some of the
theoretical aspects of graphene physics are reviewed qualitatively
by Castro Neto {\it et al.},\cite{castro} by Katsnelson,
\cite{katsnelson} and by Geim and  MacDonald; \cite{Geim07} a more
comprehensive review is given by Castro Neto {\it et al.}.\cite{rmp}
For a review on Klein tunneling see the work by
Beenakker.\cite{rmpBeenakker}


\section{Basic definitions.}

As described in the previous section, we assume that it is possible
to manufacture slabs with SiO$_2$-BN interfaces, on top of which a graphene
flake is deposit. This will induce spatial regions where graphene has
a vanishing gap intercalated with regions where the BN will cause a
finite gap.
 
\begin{figure}
    \begin{center}
     \includegraphics [scale=.55]{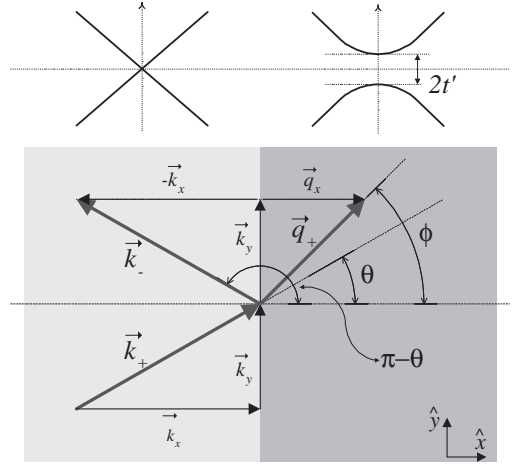}%
    \end{center}
    \caption{Top figure: graphene band structure for the massless and 
massive
    cases. In this latter, the quasi-parabolic bands have a gap energy $2t'$.
      Bottom figure: geometry in the reflection in the $k-q$ interface.
      An incident wavefunction $\psi_k^+$ with wavevector
      $\vec{k}_+$ is reflected and refracted into the $\psi_k^-$  and the $\psi_q^+$ wavefunctions
      with wavevectors
      $\vec{k}_-$ and
      $\vec{q}_+$, respectively. Since the momentum is conserved at the interface, one has that
      $\vec{q}_y=\vec{k}_y$. The refracted wave propagates with an angle $\phi$,
      which is slightly larger than the incident and reflected angles
      $\theta$ with $|\vec{q}_x|<|\vec{k}_x|$, a consequence of energy conservation.}
    \label{geometry}
   \end{figure}

In the following we will consider the graphene physics in two 
different regions:
  the $k-$region, where the graphene sheet is standing on top
of SiO$_2$, and
  a $q-$region, where a mass-like term is present,
caused by BN, inducing an energy gap
of value $2t'$ (for all numerical purposes we use
$t'=$0.1 {\ttfamily eV}). 
The wavefunctions in these two regions will be referred
  by $\psi_{\vec{k}}$ and $\psi_{\vec q}$, respectively.
The geometry of the scattering process is represented in Fig. \ref{geometry}.

  The Hamiltonian for massless Dirac electrons in graphene, around the
  $\bm K$-point in the Brillouin zone is given by
  \begin{equation}
    H_g=v_F\bm \sigma\cdot \bm p\,,
  \end{equation}
  where $\bm \sigma=(\sigma_x,\sigma_y)$, $\bm p=-i\hbar\bm\nabla$,
  $\sigma_{i}$, with $i=x,y,z$, is the $i$ Pauli matrix,
  and $v_F=3ta/(2\hbar)$, with $t$ the nearest neighbor hopping matrix
  in graphene and $a$ the carbon-carbon distance. 
Therefore, in
the massless wave function, in the $k-$region ($t'=0$), is given by 
 \begin{eqnarray}
   \psi_{\vec{k},s}={1\over\sqrt{2}}
\left(
\begin{array}{c}
{1}\\
{u(\vec{k},s)}
\end{array}
\right)
\,e^{i\vec{k}\cdot\vec{r}},
   \label{psi_k}
 \end{eqnarray}
 with
 \begin{eqnarray}
   u(\vec{k},s)=s\,e^{i\theta},
   \label{def_u}
 \end{eqnarray}
 $s=\texttt{sign}(E)$ and
 $\theta=\arctan{(k_y/k_x)}$. The corresponding energy eingenvalue is
 \begin{eqnarray}
   E=\pm v_F\hbar \sqrt{k_x^2+k_y^2}=\pm\hbar v_Fk,
   \label{free_energy}
 \end{eqnarray}
 with $k$ the absolute value of the wavevector.

In a region of finite mass the Hamiltonian for 
Dirac electrons is
\begin{equation}
    H_g=v_F\bm \sigma\cdot\bm p +t'\sigma_z\,,
\label{Hmass}
  \end{equation}
with $mv_F^2=t'$ the mass term ($m$ is the effective mass); 
as a consequence the electronic spectrum
will present a finite energy gap of value $2t'$.
 In the $q$-region (the gaped region, $t'\ne 0$), the wave funcion is
  \begin{eqnarray}
   \psi_{\vec{q},s}={1\over\sqrt{2}}
\left(
\begin{array}{c}
{1}\\
{v(\vec{q},s)}
\end{array}
\right)
\,e^{i\vec{q}\cdot\vec{r}},
   \label{psi_q}
 \end{eqnarray}
 where 
  \begin{eqnarray}
    v(\vec{q},s)=\frac{E-t'}{\hbar v_F(q_x-iq_y)}.
    \label{def_v}
  \end{eqnarray}
 Due to momentum conservation, electrons propagating through a $k-q$  
interface will conserve
 their wavevector component parallel to the interface. Thus, taking
 this interface to be located along the $\hat y$ axis, 
we will have always $k_y=q_y$.
 The $q$-region eigenenergy, associated with the eigenstate
(\ref{psi_q}) and the Hamiltonian (\ref{Hmass}), is
 \begin{eqnarray}
   E=\pm\sqrt{(q_x^2+k_y^2)(\hbar\,v_F)^2+t'^2}.
   \label{gap_energy}
 \end{eqnarray}
It is amusing to notice that the spectrum (\ref{gap_energy})
has the same form as for the electrons in a graphene bilayer,
when the two graphene planes are at different electrostatic potentials.
\cite{McCann,Pereira}
 Using Eqs. (\ref{gap_energy}) and (\ref{free_energy}), we write
 \begin{eqnarray}
   v_F\hbar q_x=\sqrt{E^2\cos^2{(\theta)}-t'^2}
 \end{eqnarray}
 and, depending on $E^2\cos^2{(\theta)}$ being larger or smaller than
 $t'^2$, $q_x$ may take a real or a pure imaginary value. Wave
 propagation follows for the former case, evanescent waves in the
 latter.

 For a real $q_x$, and since $q_y=k_y$, we have
 \begin{eqnarray}
   \sqrt{E^2-t'^2}\sin{(\phi)}=|E|\sin{(\theta)}
   \label{snell}
 \end{eqnarray}
 where $\phi$ is the angle of propagation of the electron in the
 $q$-region (see Fig.\ref{geometry}). Equation (\ref{snell}) is just the usual Snell's law, for electrons being
 refracted at the interface separating the $k-$ and $q-$ regions. We see that
 $\phi\geq\theta$ whenever $|E|>t'$.

 \subsection{Forward and backward propagation.}
 We consider now the simple reflection in the interface, with the incident and the reflected waves
 both on the $k-$ or on the $q-$region.

 In the $k-$region, the $\hat{x}-$component of the wavevector
 of the reflected wave is symmetrical with respect to the incident
 wave. Thus, for this case, we have the following transformations under a
 reflection (see also Fig.\ref{geometry})
 \begin{eqnarray}
    k_x\rightarrow -k_x~~~\rm{and}~~~e^{i\theta}\rightarrow
    e^{i(\pi-\theta)}=-e^{-i\theta}.
    \label{kx_to_minus_kx}
 \end{eqnarray}
 This leads to the generalization of Eqs.
 (\ref{psi_k}) and  (\ref{def_u}) 
  \begin{eqnarray}
   u_\pm=\pm s\,e^{\pm i\theta},
   \label{def_u_pm}
  \end{eqnarray}
  \begin{eqnarray}
   \psi^{\pm}_{\vec{k}}=\frac{1}{\sqrt{2}}
\left(
\begin{array}{c}
{1}\\
{u_\pm}
\end{array}
\right)
\,e^{\pm ik_x x+ik_y y}.
   \label{psi_k_pm}
  \end{eqnarray}
 where the plus and minus signs refers to waves propagating, respectively, in the
 positive and negative directions of the $\hat x$-axis.

 A similar reasoning leads to the generalization of (\ref{def_v}) to
 $v(\vec{q},s)$,
 \begin{eqnarray}
    v_\pm=\frac{E-t'}{\hbar v_F(\pm q_x-ik_y)}.
    \label{def_v_pm}
  \end{eqnarray}
 and, also, of the $q-$region wavefunction to
 \begin{eqnarray}
   \psi^{\pm}_{\vec{q}}={1\over\sqrt{2}}
\left(
\begin{array}{c}
{1}\\
{v_\pm}
\end{array}
\right)
\,e^{\pm iq_x x+ik_y y}.
   \label{psi_q_pm}
 \end{eqnarray}

 The differences we have just highlighted on the wavefunctions and coefficients for
 forward and backward propagating particles can be also seen in
the 
 differences in positive and negative angles of incidence in the
 interface. This changes are useful, when a guiding-wave kind of device
is made. Let us therefore analyze the case when $k_y\rightarrow -k_y$.
If in Eq. (\ref{kx_to_minus_kx}) we keep $k_x$ unaltered
 and ``reflect'' instead $k_y$ we would obtain
  \begin{eqnarray}
    k_y\rightarrow -k_y~~~\rm{and}~~~e^{i\theta}\rightarrow
    e^{-i\theta},
    \label{ky_to_minus_ky}
 \end{eqnarray}
 with similar relations for $\phi$, the angle in the q-region. For this cases, we get
 \begin{eqnarray}
     v_\pm(-\phi)=-v_\mp(\phi~)~~\rm{and}~~u_\pm(-\theta)=-u_\mp(\theta).
    \label{def_by_angle}
 \end{eqnarray}
 Apart from a minus sign, these relations shows that the
 the operation of changing the sign of $k_y$ (i.e., the angle of incidence)
 is equivalent to the one of changing the sign of $k_x$ ($q_x$ in the
 q-region). Of curse that the extra minus sign in the right hand side of 
both expressions in Eq.
 (\ref{def_by_angle}) are of no consequence within the 
calculation of reflection 
and transmission coefficients that follow.


 \subsection{Real and evanescent waves in the $q$-region}
  Since in the $q$-region there is a gap in the energy spectrum  then $q_x$
 can take both real and pure imaginary values. In the first case, we
 have wave propagation in this region, in the latter just evanescent
 waves. No simple expression as the one given by Eq. (\ref{def_u_pm}) can be
 written in this case. 
Instead, we need to consider separately the cases where $q_x$
 is a real or a pure imaginary number.

 \subsubsection{For $q_x$ real.}

 For real $q_x$, we can write a similar expression to the one in 
Eq. (\ref{def_u_pm}),
 \begin{eqnarray}
    v_\pm=\pm ve^{\pm i\phi},
    \label{def_v_free}
 \end{eqnarray}
with $\phi$ given by Eq. (\ref{snell}) and
  \begin{eqnarray}
    v=\frac{E-t'}{v_F\hbar|q|}=\frac{E-t'}{\sqrt{E^2-t'^2}},
    \label{def_v_real}
  \end{eqnarray}
where Eq. (\ref{gap_energy}) was used.
\subsubsection{For $q_x$ pure imaginary.}

 Since $k_y$ is always a real number, Eq. (\ref{def_v_pm}) implies that, if $q_x$ is a pure-imaginary,
 $v_\pm$ also is, and then
 \begin{eqnarray}
    v_\pm=\mp i\nu_\pm,
    \label{def_v_gap}
 \end{eqnarray}
 where
 \begin{eqnarray}
   \nu_\pm=\pm\frac{E-t'}{\hbar v_F(\pm k_y-\alpha)},
 \end{eqnarray}
 with the real {\it absorption coefficient} $\alpha$ defined as
 $q_x=i\alpha$, and $\alpha$ given by
 \begin{eqnarray}
  \alpha=(v_F\hbar)^{-1}\sqrt{t'^2-E^2\cos(\theta)^2}\,.
  \label{alpha}
 \end{eqnarray}

\subsubsection{Complex conjugate of the $u_\pm$ and $v_\pm$ coefficients.}

 For the calculation of the intensity reflection and transmission 
coefficient
we
 will need to deal with the complex conjugate of the $u_\pm$ and $v_\pm$ coefficients.
 The definition (\ref{def_u_pm}) for $u_\pm$ implies that
 \begin{eqnarray}
  u_\pm^*=-u_\mp.
  \label{u_alg}
 \end{eqnarray}

 In the case of $v_\pm$, its complex conjugate depends on the the fact of having a real or imaginary $q_x$,
  \begin{eqnarray}
   v_\pm^*=\left\{
     \begin{array}{cl}
       -v_\mp & ~~~{\rm if}~q_x~{\rm is~~real} \\
       -v_\pm & ~~~{\rm if}~q_x~{\rm is~~imaginary} \\
      \end{array}\right.
   \label{v_alg}
  \end{eqnarray}

 \section{Transmission and reflection at the interface: the step case.}
 \subsection{Reflection and transmission amplitude coefficients.}

\begin{figure}
   \begin{center}
    \includegraphics [scale=.75]{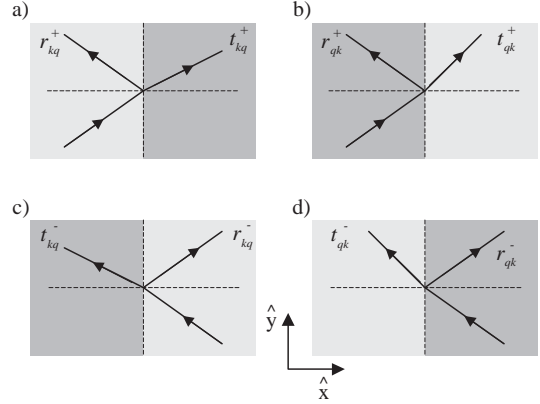}%
   \end{center}
   \caption{The four different possible cases for reflection/transmission in an interface between
    $k$ and $q$ regions.}
   \label{fig1}
  \end{figure}

 We compute now the reflection and transmission amplitude
 coefficients for electrons crossing an interface between a $k-$ and
 a $q-$ region. Unlike what happens in optics and due to the
 differences on back and forward propagation, we will need to
 consider not only two but four different cases: electrons crossing the interface
 coming from the $k-$region in the forward and backward senses, and those crossing the interface
 coming from the $q-$region, also propagating on the positive and negative senses of the $x$ axis.
 These four cases are summarized in Fig. \ref{fig1}.

 \subsubsection{Propagation from a $k-$ into a $q-$region.}

 We start by deriving the amplitude reflection and transmission coefficients, which will be denoted as
 $r^\pm_{kq}$ and $t^\pm_{kq}$ respectively, for the case of the propagation from a $k-$ into a $q-$ region.
 This situation is described in Fig. \ref{geometry} and also in  
Fig. \ref{fig1}.a).

 Since there is a partially reflected wave, the total wave function in the $k-$region must be written as
 an superposition of one associated with the incident electrons and other with those that are reflected,
 \begin{eqnarray}
   \Psi_k(\vec r)=A\,\psi^{+}_{\vec{k}}+B\,\psi^{-}_{\vec{k}}.
\label{psi_kA} 
\end{eqnarray}
 $A$ and $B$ are the normalized amplitudes for the incident and
 reflected wave functions. In the $q$-region, with $C$ the amplitude of the transmitted wave function, we have
 \begin{eqnarray}
   \Psi_q(\vec r)=C\,\psi^+_{\vec{q}}.
\label{psi_qC}
 \end{eqnarray}
 Using Eqs. (\ref{psi_kA}) and (\ref{psi_qC}), and imposing
 the continuity condition of the particle's wave function at the interface, i.e.
 $\Psi_k(x,y=0)=\Psi_q(x,y=0)$, we
 find
\begin{eqnarray}
  r^+_{kq}={B\over A}=\frac{v_+-u_+}{u_--v_+}~~\rm{and}~~t^+_{kq}={C\over A}=1+r^+_{kq},
  \label{rp_kq}
\end{eqnarray}
where the superscript $+$ recall that the incident wave function is,
in this case, traveling in the positive direction of the
$x$-axis.

 Had we considered the case were the particles travel in the backward
 direction, represented by  Fig.\ref{fig1}.c), we would have obtained
\begin{eqnarray}
  r^-_{kq}=\frac{v_--u_-}{u_+-v_-}~~{\rm and}~~t^-_{kq}=1+r^-_{kq}.
  \label{rm_kq}
\end{eqnarray}
This result can be obtain simply by exchanging the plus by the minus
signs in Eq. (\ref{rp_kq}).


 \subsubsection{Propagation from the $q-$ into the $k-$region.}

 For computing the reflection and transmission coefficients for the cases where the electrons
 come from the $q-$region into the $k$-region, $r^\pm_{qk}$ and
 $t^\pm_{qk}$, we need only to exchange $u\leftrightarrow v$ in the corresponding
 backward and forward expressions (\ref{rp_kq}) and (\ref{rm_kq}).
 The result is
 \begin{equation}
   r^\pm_{qk}=\frac{u_\pm-v_\pm}{v_\mp-u_\pm}~~{\rm and}~~
t^\pm_{qk}=1+r^\pm_{qk}.
   \label{rpm_qk}
 \end{equation}

 \subsection{Amplitude coefficients: general algebraic relations.}

  It will be very useful in the following, for expression simplification 
purposes,
  the derivation of simple relations between the reflection and
  transmission amplitude coefficients. Similar relations to those we
  present here exist also for the photons' optics case. For instance, we may write
  $
    \pm r_{12}+t_{12}=1
  $
  when a light beam is reflected and refracted in a diopter between
  regions $1$ and $2$, with the plus or minus sign corresponding respectively to
  the cases where $n_1$, the index of refraction of medium $1$, is
  smaller or larger that the one of region $2$.

  Here, however, we have that in general $r^+_{mn}\neq r^-_{mn}$ (and similarly for the transmission coefficients)
  and these relations are less trivial 
(we have used the notation $m\neq n=\{k,q\}$).

  Using the definitions in Eqs. (\ref{rp_kq}), (\ref{rm_kq}) and
  (\ref{rpm_qk}), we can write
  \begin{eqnarray}
    \mathcal{R}+\mathcal{T}=1,
    \label{RR_plus_TT_1}
  \end{eqnarray}
  where
  \begin{eqnarray}
  \begin{array}{c}
    \mathcal{R}=r^+_{kq}r^-_{kq}= r^+_{qk}r^-_{qk} \\
    \mathcal{T}=t^+_{qk}t^+_{kq}= t^-_{qk}t^-_{kq}. \\
  \end{array}
    \label{RR_TT_def}
  \end{eqnarray}
  These relations are general and don't depend on the $q_x$ being real or imaginary.
  Another general relations, useful to simplify
  expressions of the transmission of multi-layered structures, are
  \begin{eqnarray}
    r^\pm_{mn}r^\mp_{nm}=- \mathcal{R}\times{t^\mp_{nm}\over t^\pm_{nm}}\,.
    \label{odd_rel}
  \end{eqnarray}

  \subsection{Intensity reflection and transmission coefficients.}

 The general definitions for the intensity reflection and transmission coefficients are
  \begin{eqnarray}
   \begin{array}{lll}
    R^\pm_{mn}=&r^\pm_{mn}\left(r^\pm_{mn}\right)^*&\\
    T^\pm_{mn}=&t^\pm_{mn}\left(t^\pm_{mn}\right)^*=&1-R^\pm_{mn},\\
   \end{array}
   \label{RT_def}
  \end{eqnarray}
 where we keep the same notation as before. We will consider now,
 separately, the cases were $q_x$ is a real number or a pure imaginary.
 
\subsubsection{For $q_x$ real.}

 For $q_x$ a real number, we note first that for any $m\neq n=\{k,q\}$,
  \begin{eqnarray}
   \left\{\begin{array}{c}
           \left(r^\pm_{mn}\right)^*=r^\mp_{mn} \\
            \\
           \mathcal{R}=\mathcal{R}^* \\
         \end{array}
   \right.
   \label{r_rel_out_gap}
  \end{eqnarray}
These relations are just a consequence of the definitions of
$r^\pm_{mn}$ in Eqs. (\ref{rp_kq}), (\ref{rm_kq}), and (\ref{rpm_qk}),
 and in Eqs.
(\ref{u_alg}) and (\ref{v_alg}). Using Eq. (\ref{r_rel_out_gap}) in
Eq. (\ref{RT_def}) results in 
  $$
    R=\mathcal{R}~~~{\rm and}~~~T=\mathcal{T}~~~{\rm for}~q_x~{\rm real},
  $$
 which is valid, for both interfaces and both directions of propagation. Furthermore, using Eq. (\ref{RR_plus_TT_1})
 we get
 $$
   R+T=1,
 $$
 an expected result.

 Explicitly, the $\mathcal{R}$ coefficient defined in Eq. (\ref{RR_TT_def})
 is given by
  $$
    \mathcal{R}=\frac{(v_+v_--1)-(u_-v_++u_+v_-)}{(v_+v_--1)-(u_-v_-+u_+v_+)}.
  $$
 Making use of Eqs. (\ref{def_v_pm}) and (\ref{def_u_pm}), we may write
  \begin{eqnarray}
    v_+v_-&=&-v^2,~~
    u_\pm v_\mp=-s\,v\,e^{\pm
    i(\theta-\phi)}\nonumber\\
    &&{\rm and}~~u_\pm v_\pm=s\,v\,e^{\pm i(\theta+\phi)},\nonumber
  \end{eqnarray}
 where $v$ is given by Eq. (\ref{def_v}) and $\phi$ by Eq. (\ref{snell}).
 Using these expressions we obtain
  $$
    \mathcal{R}=\frac{1+v^2-2\,s\,v\cos{(\theta-\phi)}}
                     {1+v^2+2\,s\,v\cos{(\theta+\phi)}},
  $$
  where
  $$
    v\cos{(\theta\pm\phi)}= \frac{v_F\hbar}{E+t'}\left(q_x\cos{(\theta)}\mp
    k_y\sin{(\theta)}\right).
  $$
  Finally, after algebraic simplification, we obtain
  \begin{eqnarray}
   R=\mathcal{R}=\frac{k_x-q_x}{k_x+q_x}.
   \label{R_real}
  \end{eqnarray}

 \subsubsection{For $q_x$ pure imaginary.}

 For $q_x$ a pure imaginary, we see that
  \begin{eqnarray}
  \left\{\begin{array}{c}
      \left(r^\pm_{mn}\right)^*r^\pm_{mn}=1\\
      \\
      \mathcal{R}\,\mathcal{R}^* =1\\
         \end{array}
   \right.
   \label{r_rel_in_gap}
  \end{eqnarray}
 Using these relations along with Eq. (\ref{RT_def}),
 we straightforwardly obtain
  \begin{eqnarray}
    R^\pm_{kq}=1 ~~{\rm and}~~
    T^\pm_{kq}=0~~{\rm with}~~q_x~{\rm a~~ pure~~ imaginary}.
  \end{eqnarray}
 This is an expected result since the transmission $T=1-R$ must be
 zero in the case where the wave in the $k$-region enters in the gap of the $q$-region.
 If the incident wave propagates in the gap region, i.e. it is an evanescent wave,
 the coefficients $R^\pm_{qk}$ and $T^\pm_{qk}$ are physically meaningless.

  We see from the second expression in (\ref{r_rel_in_gap}) that $\mathcal{R}$ is a modulo 1 complex
  quantity. It may be written as
 \begin{eqnarray}
   \mathcal{R}=e^{i2\varphi}\,,
   \label{def_varphi}
 \end{eqnarray}
with $2\varphi$ a convenient definition of its argument. To compute
this angle, in the spirit of Eq. (\ref{alpha}), we replace
$q_x$ by $i\alpha$ in Eq. (\ref{R_real}) to obtain
 $$
   \mathcal{R}=\frac{k_x-i\alpha}{k_x+i\alpha}.
 $$
 Computing the real part of this quantity, we get
 $$
   \cos{(2\varphi)}=\frac{2E^2\cos(\theta)^2}{t'^2}-1
 $$
 and, after straightforward manipulation,
 \begin{eqnarray}
  \cos{(\varphi)}\,t'=\cos{(\theta)}\,|E|~~{\rm or~~ else}~~
      \tan{(\varphi)}={\alpha\over k_x},
  \label{ev_snell}
 \end{eqnarray}
 a Snell type expression for the $q_x$ pure imaginary case.

\begin{figure}
    \begin{center}
     \includegraphics [scale=.55]{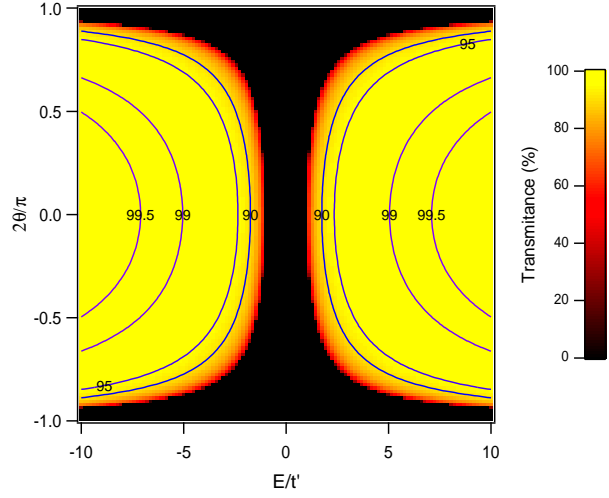}
    \end{center}
    \caption{Intensity transmission for particles crossing the interface from a $k$-region into a $q$-region.
    The black region corresponds to a zero transmission, a case that correspond to the total internal reflection in
    the usual photonic optics.}
    \label{T_step0}
   \end{figure}

 Since $R=\mathcal{R}\mathcal{R}^*=1$ for $q_x$ a pure imaginary, a general
 expression for the intensity reflection and transmissions coefficients 
(valid for $q_x$ both real and pure imaginary)
is given by
 \begin{eqnarray}
   R&=&1-T\nonumber\\
   &=&\left|\frac{k_x-q_x}{k_x+q_x}\right|\nonumber\\
   &=&\left|\frac{1+v^2-2s\,v\cos{(\theta-\phi)}}{1+v^2+2s\,v\cos{(\theta+\phi)}}\right|.
   \label{R_step}
 \end{eqnarray}
Naturally, Eq. (\ref{R_step}) depends on $t'$, since both $v$, $\theta$,
and $\phi$ depends on this quantity. When one considers the case
$\theta=\phi=t'=0$ one obtains $R=0$.
 This expression is plotted in a density/contour plot in Fig.\ref{T_step0}.

\section{The barrier.}

\begin{figure}
    \begin{center}
     \includegraphics [scale=.85]{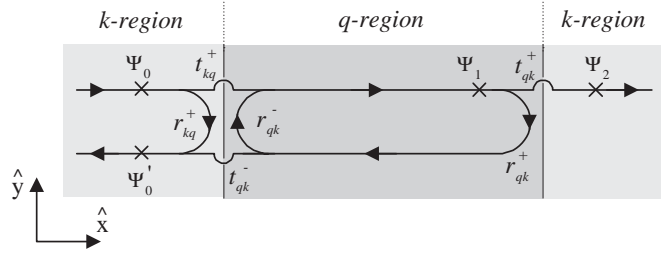}%
    \end{center}
    \caption{Barrier: scheme for the computation of the transmission.}
    \label{barrier_scheme}
   \end{figure}

 With the above definitions, the computation of transmission and reflections coefficients for any type of
 multi-interface device follows similar expressions as those found in normal
 optics.\footnote{There is no analog, however, for the gap-region with normal incidence.}
 To illustrate this, we consider in the following a heterostructure made of
 a $q-$region of width $w$ placed between two semi-infinite slabs of $k$-regions,
 as shown in the Fig. \ref{barrier_scheme}. Our goal will be the
 derivation of the intensity transmission coefficient for this case,
 which we will denoted by $T_b$. 
We notice that results for barriers of the same height when the spectrum
of the the electrons is linear in every spatial regions was considered
in Ref. \cite{Pereira2}.

 In Figure \ref{barrier_scheme}, the wave function $\Psi_1$ describes an electron, traveling in the
positive direction of $\hat{x}-$axis, just before crossing the
{\it diopter} $q-k$. This wave function can be seen as resulting
from the coherent superposition of two wave functions, one being
itself after a round trip in the $q-$region, given by
$\Psi_1\,t^+_{qk}t^-_{qk}\,e^{i2q_xw}$, and another one which is the
{\it incident} wave funtion $\Psi_0$ after crossing the first
interface $k-q$, equal to $\Psi_0\,t^+_{kq}\,e^{iq_xw}$. Adding
these two contribution and solving in order to $\Psi_1$ we obtain,
 $$
   \Psi_1=\Psi_0\frac{t^+_{kq}\,e^{iq_xw}}{1-r^+_{qk}r^-_{qk}\,e^{i2q_xw}}.
 $$
If we denote the amplitude transmission coefficient for this barrier
as $t_B=\Psi_2/\Psi_0$ and using the fact that
$\Psi_2=t^+_{qk}\Psi_1$, we finally obtain
 \begin{eqnarray}
  t_B=\frac{\mathcal{T}e^{iq_xw}}
                                 {1-\mathcal{R}e^{i2q_xw}}\,
  \label{amp_trans_coef_barrier}
 \end{eqnarray}
where the definitions (\ref{RR_plus_TT_1}) were used.


\subsection{$q_x$ real: free propagation.}

 If there is wave propagation in the $q$-region, $q_x$ is
real, $\mathcal{R}=R$ and $\mathcal{T}=T$, and
 \begin{eqnarray}
   T_B=t_B\,t_B^*=1-R_B=\left[1+\left(\frac{2}{\pi}\mathcal{F}\right)^2\sin^2{(q_xw)}\right]^{-1}.
  \label{int_trans_barrier}
 \end{eqnarray}
 where we used the finesse definition
 \begin{eqnarray}
   \mathcal{F}=\pi\frac{\sqrt{R}}{T}={\pi\over 2}\frac{t'}{q_x},
   \label{finesse}
 \end{eqnarray}
 to highlight the similarity with an Fabry-P\'erot solid etalon (made of glass, for example)
 in usual optics.\cite{BornWolf} However, this similarity is
 elusive. In the solid etalon case, in general, the finesse is almost a constant coefficient since the
 interfaces' reflectivities (e.g., in a diopter glass-air) has a small dependence on the
 energy (optical resonances are typically far from
 the visible part of the spectrum and there's no gap as in the case of the graphene with a mass term).
 In the case treated here, $\mathcal{F}$ has a strong dependence on the
 energy $E$ of the particles and, furthermore, there is also a gap
 present. We will revisit a Fabry-P\'erot type of device further in this work.

  \begin{figure}
    \begin{center}
     \includegraphics [scale=.5]{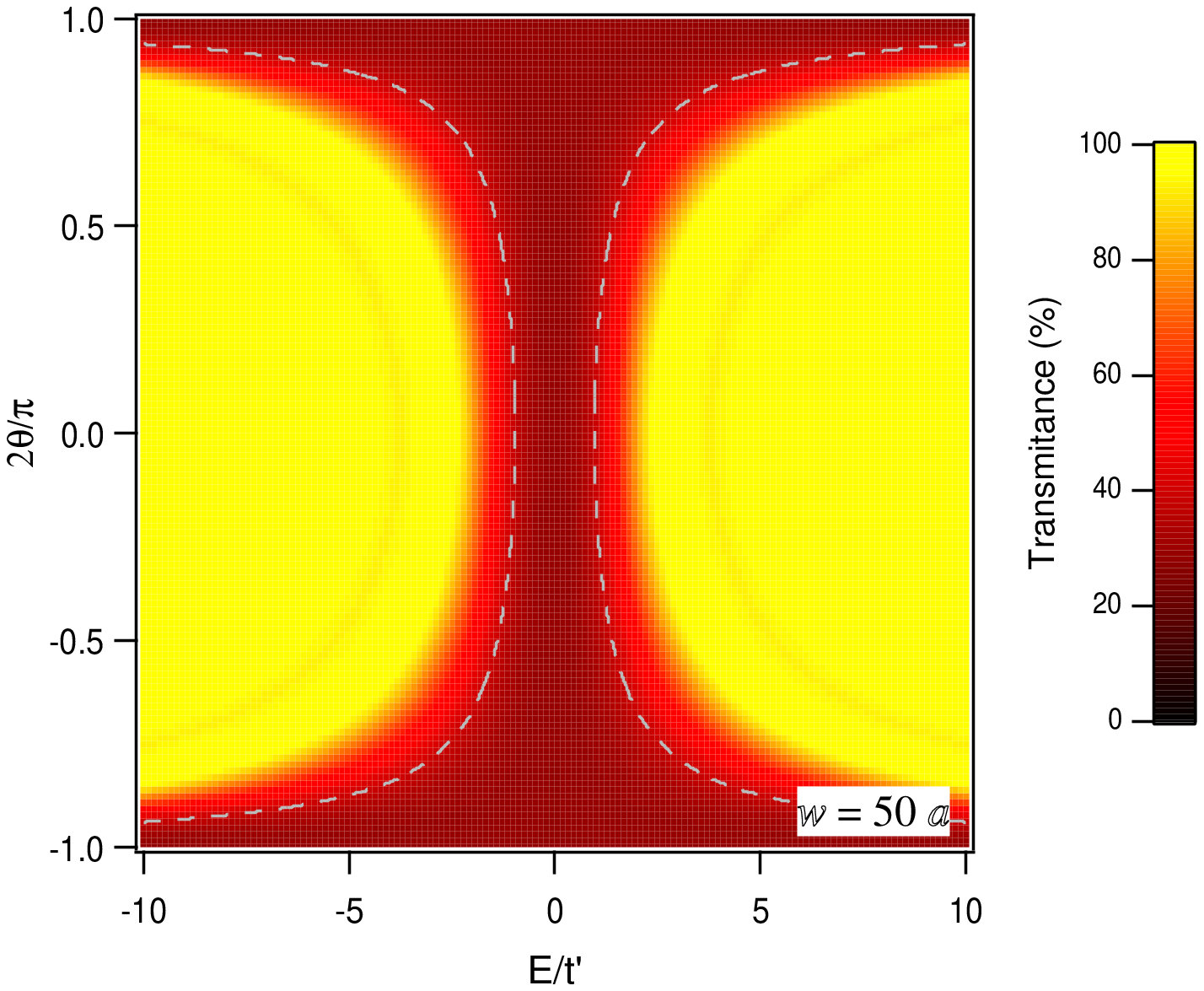}%
    \end{center}
    \begin{center}
     \includegraphics [scale=.5]{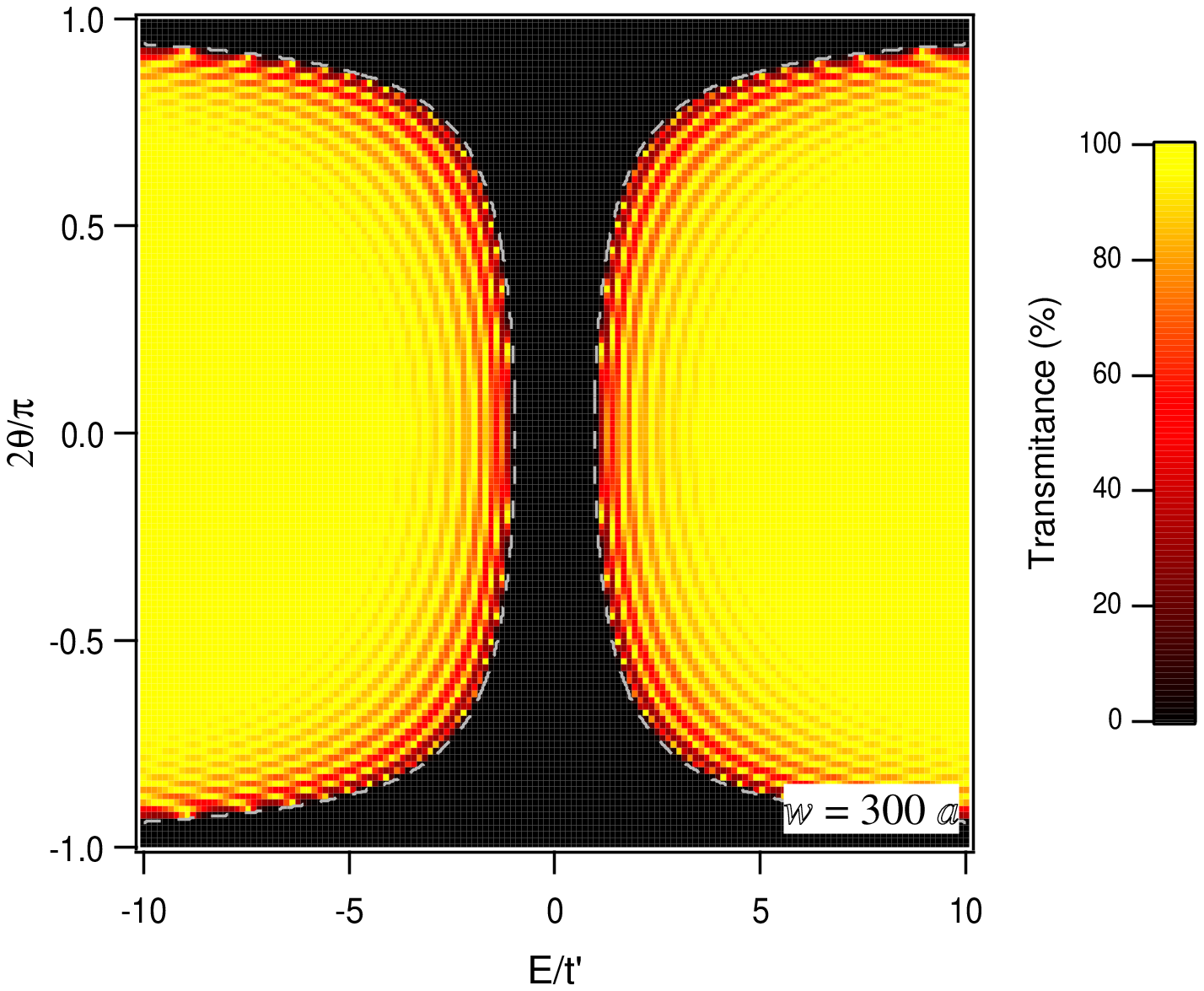}%
    \end{center}
    \caption{Transmission of a barrier for a $q$-region width of $w=50 a$ (top) and $w=300 a$ (bottom).
    For a sufficient narrow width, the wave tunnels across the $q$-region resulting in a non null transmission.
    In optics, this behavior is known as \textit{frustrated} total internal reflection.
    The dashed lines marks the region where in the step, the transmission is zero.
    }
    \label{T_step_w50_w200}
   \end{figure}

  \begin{figure}
    \begin{center}
     \includegraphics [scale=1]{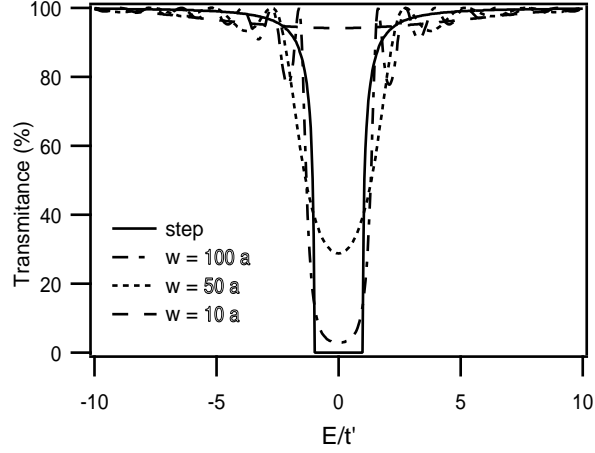}%
    \end{center}
    \caption{Transmission coefficient for a simple step and several barriers for normal incidence. In the barriers case,
    there is a \textit{frustrated} total internal reflection and, in the gap, the transmission is non zero and
    increases with decreasing values of the width of the $q$-region.}
    \label{T_step}
   \end{figure}

  \begin{figure}
    \begin{center}
     \includegraphics [scale=.8]{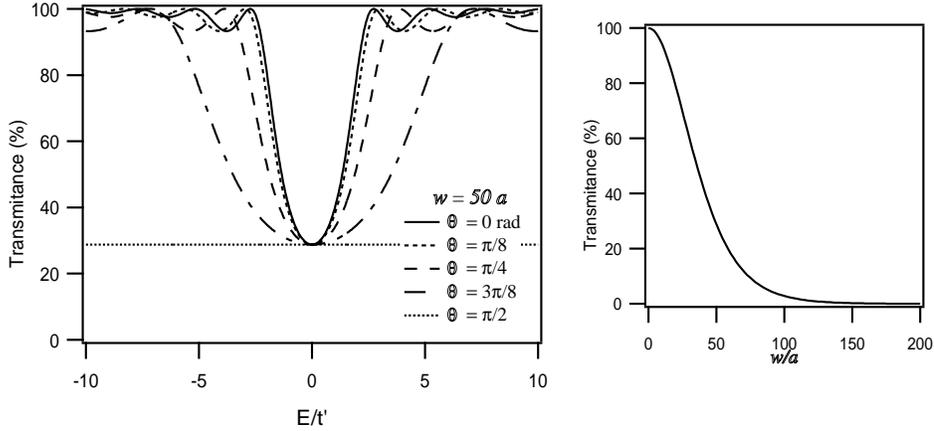}%
    \end{center}
    \caption{\label{T_barrier_w50_5thetas}
Left: 
Transmission coefficient for a barrier with a width $50a$
    and for different angles of incidence $\theta$. At $E=0$, all curves have the same value and
    for $\theta=\pi/2$, the transmission is a delta function at $E=0$.
    Right: Zero energy transmission coefficient of a barrier as 
function of its width $w$.}
   \end{figure}

 \subsection{Inside the gap: frustrated total internal reflection.}

 Inside the gap, $q_x$ is a pure imaginary and there's no wave
 propagation. This is similar to the total internal reflection in
 optics, where only an evanescent wave exists that carries no
 energy (since in here, the coefficient $R_{kq}=1$) and decays
 exponentially in the $x$ direction (although keeping the phase term
 $e^{ik_yy}$ in the $y$ direction).

 However, by placing a $k$-region nearby the evanescent wave, some of the
 energy of the totally reflected wave tunnels throughout the gap
 region, a phenomena known in optics as \textit{frustrated total
 internal reflection}. This phenomena is also described by
 Eq. (\ref{int_trans_barrier}), whenever $q_x$ becomes a pure imaginary. In this case,
 replacing $q_x=i\alpha$ and using the definitions
 (\ref{alpha}) and (\ref{def_varphi}) we may simplify
 Eq. (\ref{int_trans_barrier}) to
 \begin{eqnarray}
   T_B=t_B\,t_B^*=1-R_B
      ={1\over 1+\zeta},
  \label{int_trans_barrier_ingap}
 \end{eqnarray}
where we have used the definition
 \begin{eqnarray}
  \zeta=\frac{\sinh^2{(\alpha
      w)}}{\sin^2{(\varphi)}},
  \label{zeta}
 \end{eqnarray}
which will be used in the following.
  If $E=0$, Eq. (\ref{ev_snell}) implies that $\varphi=-\pi/2$ and
$\alpha=t'/v_F\hbar$. $T_B$ is in this case independent of $\theta$
and is equal to
\begin{eqnarray}
   T_B(E=0)=\cosh^{-2}{\left(w\frac{t'}{v_F\hbar}\right)}.
\end{eqnarray}
This behavior is clearly shown in left panel of 
Fig. \ref{T_barrier_w50_5thetas}.
In the right panel of Fig.\ref{T_barrier_w50_5thetas} it is shown how this
tunneling transmittance at zero energy varies with the barrier
width. A $50\%$ reduction is accomplished for a barrier with a width of
approximately $36~a$.

 \section{Transfer matrices}
 The method used in the last Section for computing $t_B$, although being simple becomes very
 difficult to handle for more complex hetero-structures, with more
 than two interfaces. These type of cases are usually treated with the use of 
\textit{transfer} matrices.
 These will be computed in the following.
  \begin{figure}
    \begin{center}
     \includegraphics[scale=.85]{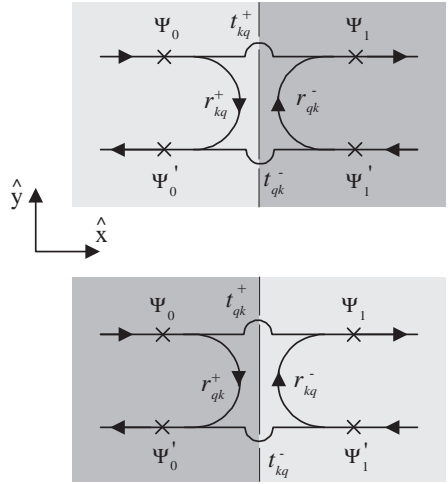}%
    \end{center}
    \caption{\label{fig2}Schemes for the computation of the transfer matrices.}
   \end{figure}
 Figure \ref{fig2} shows the scheme used to compute the
 transfer matrix in an interface $k-q$ and $q-k$. In both cases, our
 goal is to derive $\Psi_1$ and $\Psi'_1$ from the knowledge of $\Psi_0$ and
 $\Psi'_0$. Defining
 $$
 \left(
\begin{array}{c} 
{\Psi_1}
\\{\Psi'_1}
\end{array}
\right)
=\mathbf{M}_{mn}
\left(
\begin{array}{c}
{\Psi_0}\\
{\Psi'_0}
\end{array}
\right)
,
 $$
 where $\mathbf{M}_{mn}$ is the transfer matrix for the generic $m-n$
 interface, and using Eq. (\ref{RR_TT_def}), we obtain
 the result
 \begin{eqnarray}
   \mathbf{M}_{mn}=\left[
           \begin{array}{ccc}
             \displaystyle{ \frac{1       }{t^+_{nm}}}&
            &\displaystyle{ \frac{r^-_{nm}}{t^-_{nm}}}\\
            \\
             \displaystyle{-\frac{r^+_{mn}}{t^-_{nm}}}&
            &\displaystyle{ \frac{1       }{t^-_{nm}}}\\
           \end{array}
          \right]
 \end{eqnarray}
 The determinant of this matrix is given by
 \begin{eqnarray}
  Det(\mathbf{M}_{kq})&=&\left[Det(\mathbf{M}_{qk})\right]^{-1}=\frac{t^+_{kq}}{t^-_{qk}}\,.
 \end{eqnarray}
 As expected, $Det(\mathbf{M}_{kq}\times\mathbf{M}_{qk})=1$.
 The free propagation of a particle in a $k-$ and in a $q-$region of width $\xi$ is given, respectively, by
 \begin{eqnarray}
 \begin{array}{ll}
   \mathbf{L}_k(\xi)=\left[
           \begin{array}{cc}
             e^{ik_x\xi}&0\\
             0&e^{-ik_x\xi}\\
           \end{array}
          \right];
          &
   \mathbf{L}_q(\xi)=\left[
           \begin{array}{cc}
             e^{iq_x\xi}&0\\
             0&e^{-iq_x\xi}\\
           \end{array}
          \right]\,.
 \end{array}
 \label{free_prop_matrices}
 \end{eqnarray}

 \section{The diode.}
   \begin{figure}
    \begin{center}
     \includegraphics[scale=.7]{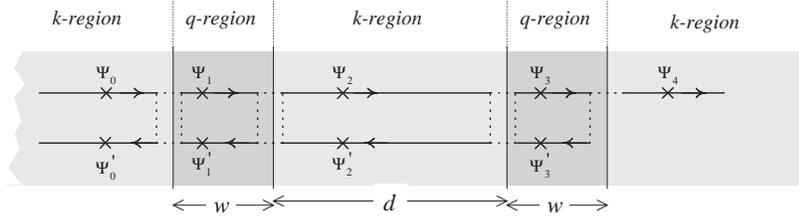}%
    \end{center}
    \caption{\label{diode}The diode heterostructure: two thin slabs of $q-$regions of width $w$ separated by a $k-$region of
           width $d$, all inside semi-infinite slabs of $k-$regions.}
   \end{figure}
 We consider now a more complex system composed by a sandwich of two $q$-regions of width $w$ separated by a slab of a
 $k-$region with width $d$, inside two semi-infinite $k-$regions. To
 derive the amplitude transmission coefficient of such a structure we need to compute the expression,
 $$
    t_d=\mathbf{M}_{qk}\,\mathbf{L}_q(w)\,\mathbf{M}_{kq}\,\mathbf{L}_k(d)\,\mathbf{M}_{qk}\,\mathbf{L}_q(w)\mathbf{M}_{kq}.
 $$
 The result of this expression can be simplified using Eq. (\ref{odd_rel}),
 resulting in
  \begin{eqnarray}
   t_D=\frac{\mathcal{T}^2\,e^{2iq_xw}}
              {(\mathcal{R}\,e^{2iq_xw}-1)^2-\mathcal{R}(e^{2iq_xw}-1)^2\,e^{2ik_xd}}\nonumber.
  \end{eqnarray}
   \begin{figure}
    \begin{center}
     \includegraphics[scale=.5]{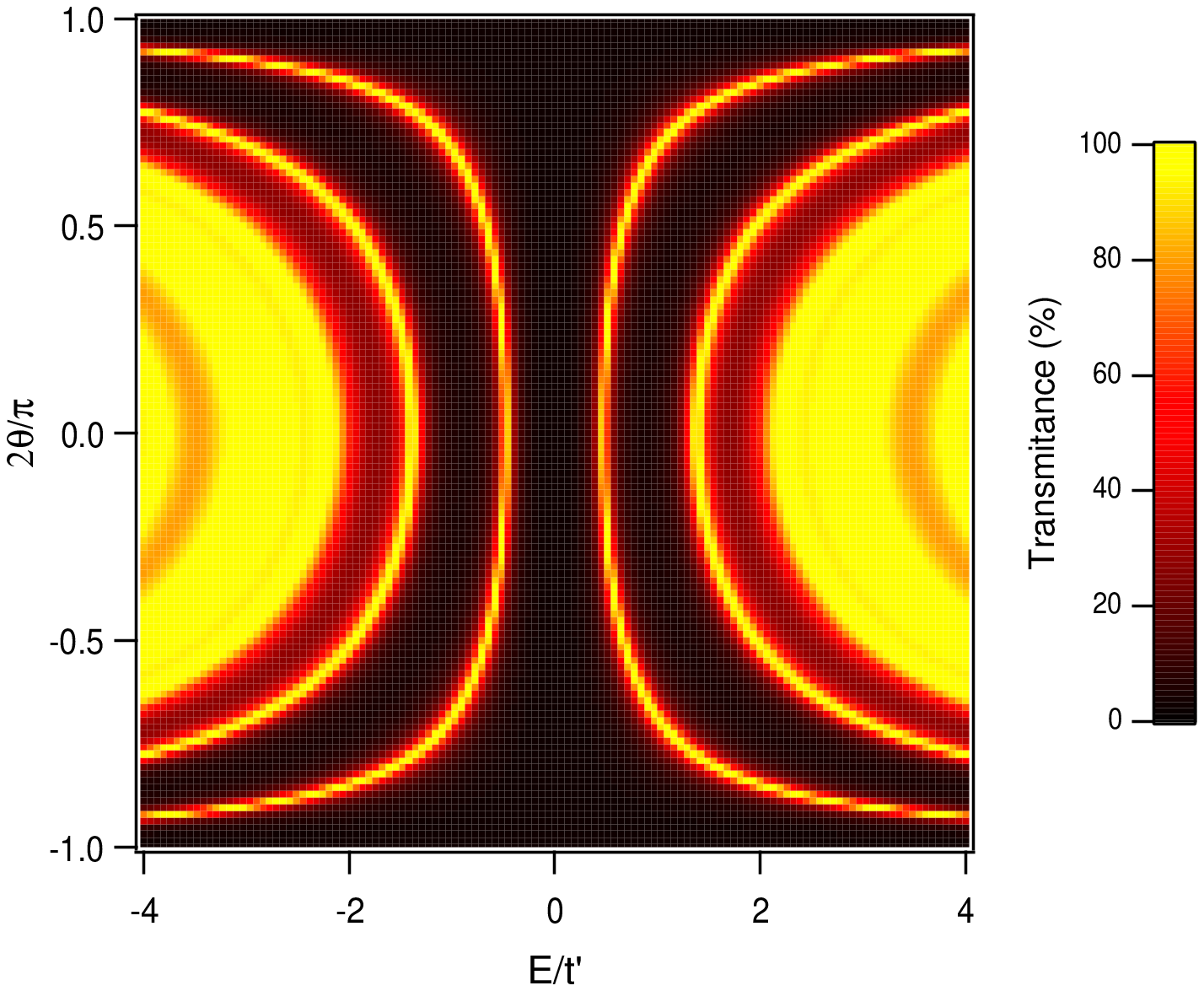}%
    \end{center}
    \begin{center}
     \includegraphics[scale=.5]{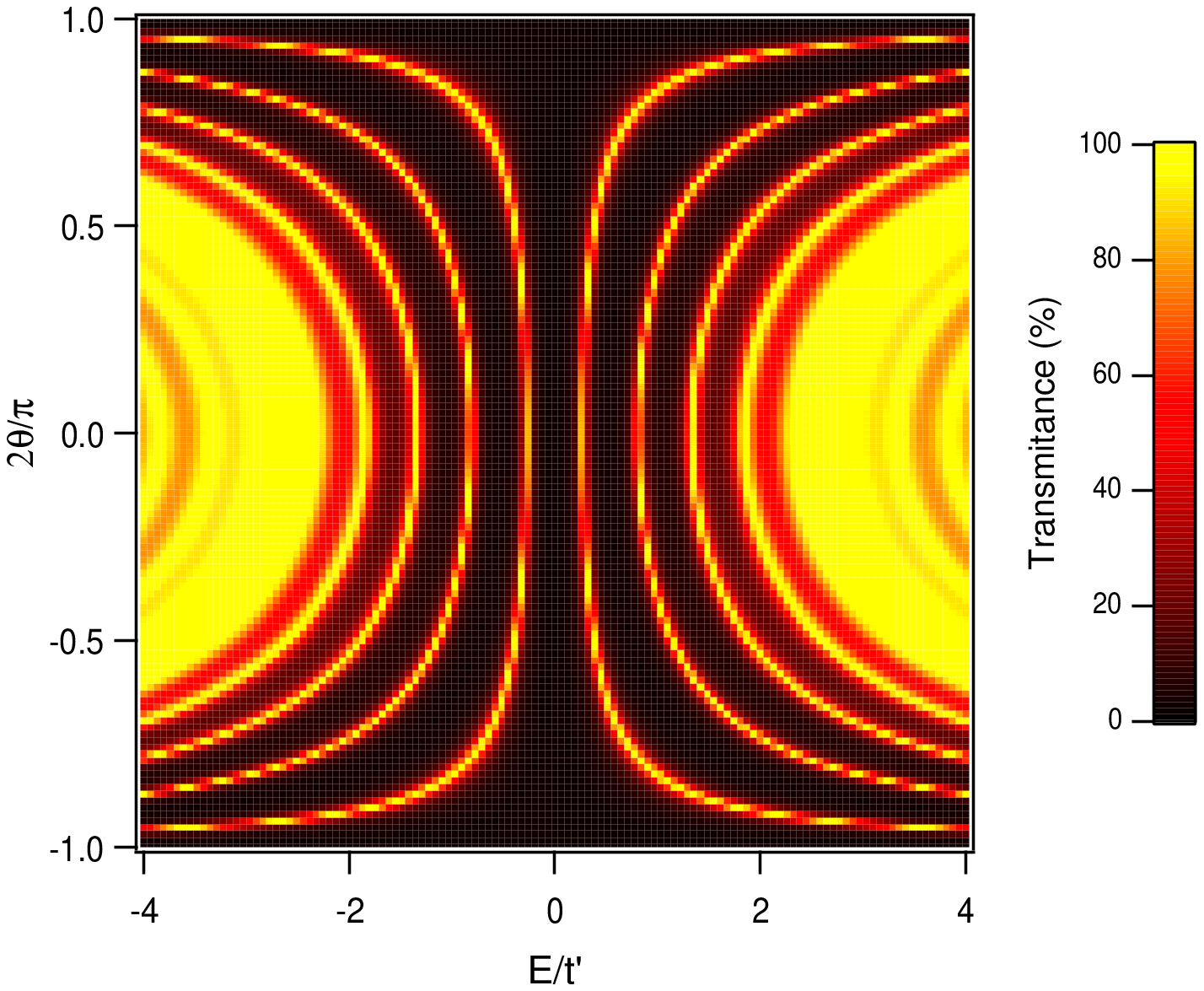}%
    \end{center}
    \label{T_diode_w50_d100_200}
    \caption{Transmission of a diode structure with $w=50a$ and $d=100a$ (top) and $d=200a$ (bottom).}
   \end{figure}
 For the most important case where the $q-$regions are barriers with
 energy higher than the energy of the particles, we have a resonant
 diode. In this case and using the definitions (\ref{alpha}) and
 (\ref{def_varphi}) we get
  \begin{eqnarray}
    t_D=t_B^2\times\left[1-\frac{\sinh^2{(\alpha w)}}{\sinh^2{(\alpha w+i\varphi)}}\,
          e^{i2k_xd}\right]^{-1},
    \label{t_D_1}
 \end{eqnarray}
 where $t_B$ is the amplitude transmission for a simple barrier,
 given by Eq. (\ref{amp_trans_coef_barrier}).

  We may simplify Eq. (\ref{t_D_1}) by expanding the term $\sinh{(\alpha w+i\varphi)}$
 and expressing the result in a complex polar representation. Doing this and
 using the definition for $\zeta$ in Eq. (\ref{zeta}) we obtain
 $$
   \frac{\sinh^2{(\alpha w)}}{\sinh^2{(\alpha w+i\varphi)}}=
   \frac{\zeta}{1+\zeta}\exp{(i2\widetilde{\varphi})}
 $$
 with the phase term argument given by
 $$
   \widetilde{\varphi}=-\arctan{\left[\coth{(\alpha w)}\tan{(\varphi)}\right]}.
 $$
 The intensity transmission coefficient can now be easily computed,
 being equal to
 \begin{eqnarray}
   T_D=
   \frac{1}{1+\left({2\over\pi}\mathcal{F}_D\right)^2
   \sin^2{(\widetilde{\varphi}+k_xd)}},
  \label{int_trans_diode}
 \end{eqnarray}
 where now, the \textit{diode finesse} $\mathcal{F}_D$ is given by
 \begin{eqnarray}
  \mathcal{F}_D=\pi\sqrt{\zeta(1+\zeta)}.
  \label{finesse_diode}
 \end{eqnarray}
  \subsection{Revisiting the Fabry-P\'erot: etalon made with "mirrors".}
 The expression  (\ref{int_trans_diode}) results in
 the simple case of a Fabry-P\'erot etalon if:
 \begin{enumerate}
    \item $\alpha w\gg1$, which implies $\coth{(\alpha w)}\cong1$ and  $\tilde{\varphi}\cong\varphi$;
    \item $E\ll t'$, which implies that $\varphi\approx\pi/2$.
 \end{enumerate}
 With these approximations we get
 $$
   \mathcal{F}_D={\pi\over2}\sinh{(2\alpha w)}
 $$
 and then
 \begin{eqnarray}
 T_D=\frac{1}{1+\sinh^2{(2\alpha w)}\cos^2{(k_xd)}},
 \label{int_trans_diode_ingap}
 \end{eqnarray}
 an expression that can be derived from a delta function potentials
 treatment, treated in  Section \ref{rtd}.
%
%

\subsection{The diode tunneling conductance.}
\begin{figure}
  \begin{center}
   \includegraphics[scale=1]{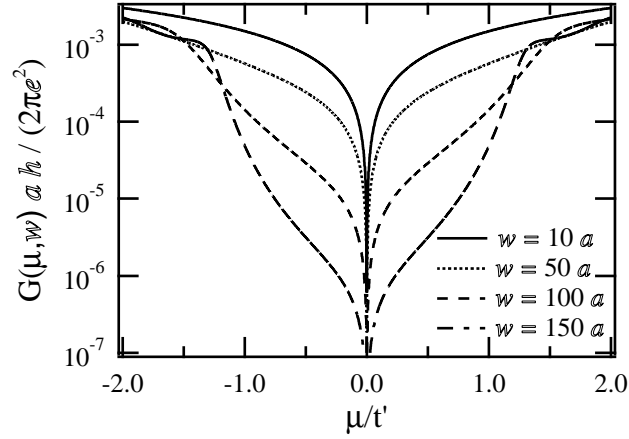}%
  \end{center}
   \caption{(color on line) Linear response conductance $G(\mu,w)$, per
    Dirac cone, as function of the chemical potential $\mu$, for
    different values of the width $w$. The hopping matrix $t$ is taken
    to be $t=2.7$ {\ttfamily eV} and $t'=$ 0.1 {\ttfamily eV}. }
\label{Fig_G_linear_response_barrier}
\end{figure}

Let us now compute the tunneling conductance of the device as
function of the potential bias $V$, the chemical potential of the
leads, and the length $w$ of the barrier. We shall assume that the
device is operating in the region where the chemical potential of
the leads lies inside the gap of the barrier. The total tunneling
current density (i. e. the current per unit of cross-sectional
length) through the device is given by
\begin{eqnarray}
J(V,w)&=&-\frac {2e}{4\pi^2v_F\hbar^2}\int d\theta
EdE\cos\theta\,T(E,\theta)
\nonumber\\
&\times& [f(E-\mu_L)-f(E-\mu_R)]
\end{eqnarray}
where $f(x)=(1+e^{x/(k_BT)})^{-1}$, $\mu_L=\mu+eV/2$,
$\mu_R=\mu-eV/2$, and $\mu$ the chemical potential of leads in the
equilibrium state. The linear response conductance per unit of
cross-sectional length is given, at zero temperature, by
\begin{equation}
G(\mu,w)=\frac {e^2}{\hbar}\frac {\vert \mu\vert }{3\pi^2a t}
\int_{-\pi/2}^{\pi/2} d\theta\cos\theta\, T(\mu,\theta). \label{G}
\end{equation}
In Fig. (\ref{Fig_G_linear_response_barrier}) we plot $G(\mu,w)$ as
function of $\mu$ for several widths $w$. It is clear that the value
of $G(\mu,w)$ may change by several orders of magnitude, close to
$\mu=0$ by a small change of $\mu$. Naturally for wider barriers one
obtains a smaller conductance.

%
%

\subsection{A limiting case.}

The limiting case of barrier can be represented by a delta-function
potential, $V(x,y)=g\sigma_z\delta(x)$. (See next section for complete
discussion of delta function potentials in the Dirac equation.) It
is interesting to compute the reflected flux for both Sch\"odinger
and Dirac electrons for this potential. In the first case one
obtains
\begin{equation}
R=\frac {(2mg/\hbar^2)^2}{4k^2_F\cos^2\theta +(2mg/\hbar^2)^2 }\,,
\end{equation}
whereas for Dirac electrons the result is
\begin{equation}
R = \tanh^2[g/(\hbar v_F)]\,.
\end{equation}
It is clear that for electrons in graphene $R$ is  angular and
energy  independent. For the case $g\gg \hbar v_F$ the reflexion
tends to unity.
\section{The diode: a limiting case.}
\label{rtd}

Finally we want to discuss a limiting case of the resonant tunneling
diode made of graphene. The device is represented in Fig.
\ref{diode}. The corresponding study for Schr\"odinger electrons
was done by Tsu and Esaki.\cite{Tsu}

A limiting situation of the device described in Fig. \ref{diode}
is one where the barriers are are describe by a scalar Lorentz
potential of the form
\begin{eqnarray}
V(x,y) &=& \lim_{\epsilon\rightarrow 0}g \frac
{1}{2\epsilon}[1-\theta(\vert x\vert-\epsilon)]\sigma_z
\nonumber\\
&+& \lim_{\epsilon\rightarrow 0}g \frac
{1}{2\epsilon}[1-\theta(\vert x-d\vert-\epsilon)]\sigma_z
\nonumber\\
&=&g\sigma_z[\delta(x)+\delta(x-d)] \,. \label{pot}
\end{eqnarray}
The connection with the true barrier is made by identifying $g$
with $\alpha t'w a$, with $\alpha$ a numerical constant of dimensions inverse
of length. 
This form of
the potential is equivalent to a mass term and therefore to a gap in
the spectrum. However, given the short range nature of the
potential, its effect comes only in the boundary conditions imposed
on
 the wave function at the potential position.

The problem of Dirac electrons in delta function potentials has been
studied in the past\cite{McKellar87a,McKellar87b} and is not without
subtleties. \cite{Sutherland81,Adame89,Roy93} The subtleties can be
traced back to the problem of evaluating the integral
\begin{equation}
\int^{\epsilon}_{-\epsilon}f(x)\delta(x)dx\,, \label{delta}
\end{equation}
where $f(x)$ is a discontinuous function at $x=0$. If we try to
solve the problem of Dirac electrons with a delta function potential
using the same trick\cite{Davies} one uses for  Schr\"odinger
electrons we face the problem defined by the integral (\ref{delta}).
This is so because the wave function of Dirac electrons in a delta
function potential is discontinuous at the point where the delta
function is located. There are several strategies to overcome this
difficulty.
\cite{McKellar87a,McKellar87b,Sutherland81,Adame89,Roy93} The most
straightforward was devised by McKellar and Stephenson
\cite{McKellar87a,McKellar87b} and generalized by Dominguez-Adame
and  Maci\'a.\cite{Adame89} In short, the Dirac equation along the
$x$ direction can be can be written as
\begin{equation}
\frac {d\phi(x)}{d x}=\hat G(x)\phi(x)\,,
\end{equation}
where $\phi(x)$ is a spinor wave function. This problem can be
formally solved as
\begin{equation}
\phi(x)=T_x e^{\int_{x_0}^x\hat G(x)dx}\phi(x_0)\,,
\end{equation}
where the operator $T_x$ is the position order operator such that
\begin{equation}
T_x[ \hat G(x)\hat G(y)]=\hat G(x)\hat G(y)\theta(x-y)+ \hat
G(y)\hat G(x)\theta(y-x)\,.
\end{equation}
Since we are interested in determine the boundary conditions obeyed
by the wave function $\phi(x)$  at the delta function position we
consider the infinitesimal interval $x\in[-\epsilon,\epsilon]$
obtaining
\begin{equation}
\phi(\epsilon)=T_x e^{\int_{-\epsilon}^\epsilon\hat
G(x)dx}\phi(-\epsilon)\,.
\end{equation}
The integral is dominated by the delta function and for the problem
we are treating in this paper we obtain the following boundary
condition
\begin{equation}
\phi(\epsilon)=e^{-i\frac
{g}{v_F\hbar}\sigma_x\sigma_z}\phi(-\epsilon)\,.
\end{equation}
To evaluate how the exponential acts on $\phi(-\epsilon)$ we use the
Lagrange-Silvester formula\cite{Adame89} for a function $f(\bm M)$
of a matrix $\bm M$
\begin{equation}
f(\bm M) = f(\lambda_1)\frac {\bm 1\lambda_2 -{\bm
M}}{\lambda_2-\lambda_1} +f(\lambda_2)\frac {\bm 1\lambda_1 -{\bm
M}}{\lambda_1-\lambda_2}\,, \label{lagrange}
\end{equation}
where $\lambda_{1,2}$ are the eigenvalues of $\bm M$. For the
problem at hand, Eq. (\ref{lagrange}) leads to the following
boundary condition around $x=0$
\begin{equation}
\left(
\begin{array}{c}
\phi_a(0^+)\\
\phi_b(0^+)
\end{array}
\right) =\cosh\tilde g \left(
\begin{array}{c}
\phi_a(0^-)\\
\phi_b(0^-)
\end{array}
\right) +i\sinh\tilde g \left(
\begin{array}{c}
\phi_b(0^-)\\
-\phi_a(0^-)
\end{array}
\right)\,,
\end{equation}
where $\tilde g =\frac {g}{v_F\hbar}$ is an adimensional interaction
constant and $0^{\pm}$ represent positive and negative
infinitesimals. A similar boundary condition holds for $x=d$. For
the potential (\ref{pot}) we can now define three different regions,
I, II, and III, defined as $x<0$, $0<x<d$, and $x>d$, respectively.
In each of these regions the wave function is a sum of two plane
waves of opposite momentum along the $x-$direction, with each plane
wave multiplied by the  coefficients $A_\Gamma$ and $B_\Gamma$,
where $\Gamma=$I, II, and III labels the three regions defined above.

\begin{figure}
 \begin{center}
   \includegraphics*[scale=.5]{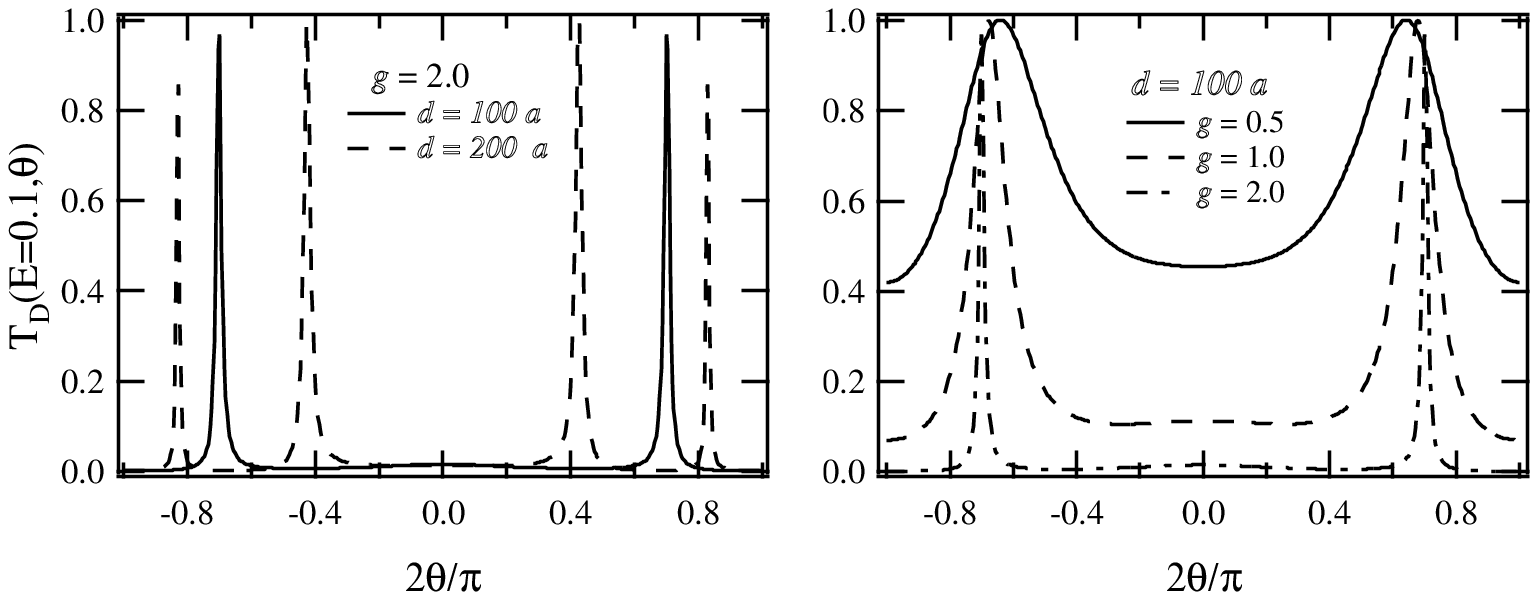}
    \includegraphics*[scale=.5]{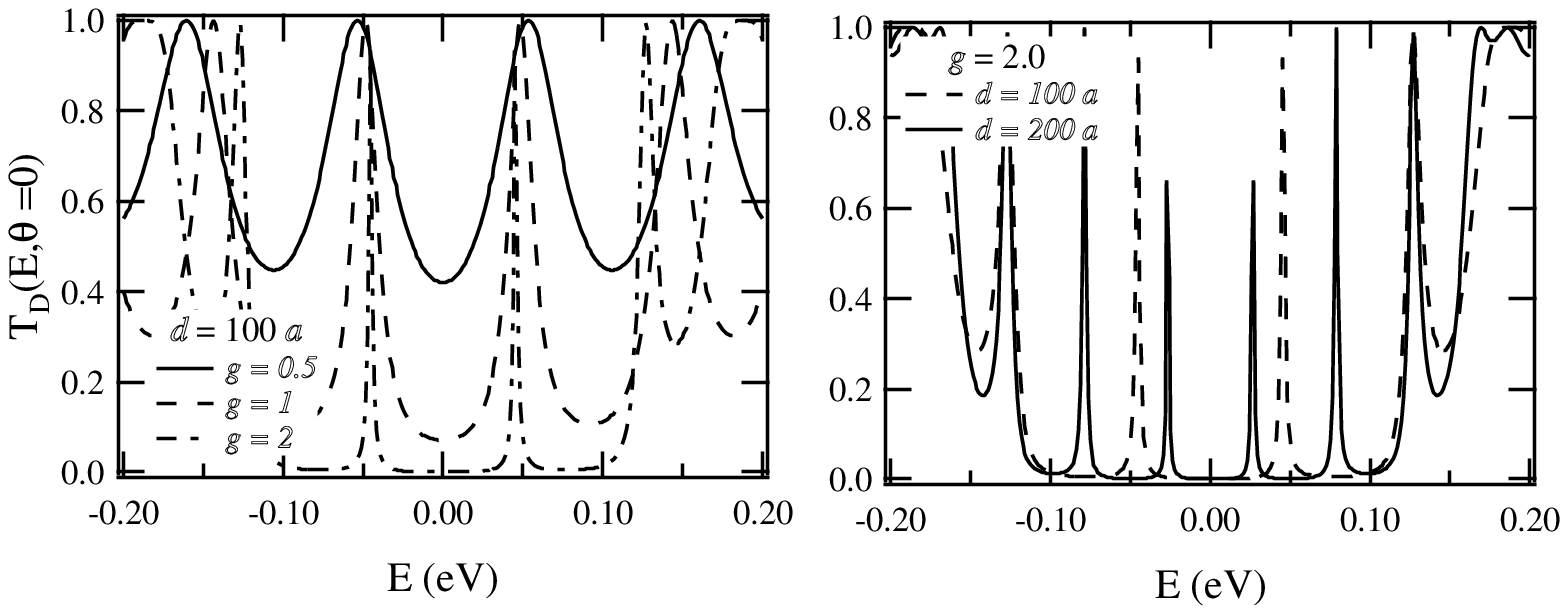}
 \end{center}
\caption{(color on line) Top panels: transmitted flux
$T_f(E=0.1,\theta)$ for fixed $\tilde g=2$ and two $d=$100$a$, 200$a$ 
and for fixed $d=100a$ and three $\tilde g=$0.5, 1, 2. 
Bottom panels: transmitted
flux $T_f(E,\theta=0)$ for the same previous cases. }
\label{Fig_RTD_transmitted_flux}
\end{figure}

Once the matrix $T$ has been computed (see appendix \ref{app})
the reflection coefficient is
obtained from
\begin{equation}
r= -\frac {T^\ast_{12}}{T^\ast_{11}}\,,
\end{equation}
and the transmitted flux $T_f$ from
\begin{equation}
T_f=1-rr^\ast\,.
\end{equation}
For the case of zero electrostatic potentials, $U_i=0$, we obtain
\begin{equation}
T_f = \frac {1} {1+\sinh^2{(2\tilde g)}\cos^2{(2dk)}}\,,
\label{tf}
\end{equation}
which is a similar expression to the one in
(\ref{int_trans_diode_ingap}) if $\tilde g=\alpha w$. It is simple to
identify the limit $T\rightarrow 1$, which occurs when
$2kd=(2n+1)\pi$, with $n=0,1,2,\ldots$.

In Figure \ref{Fig_RTD_transmitted_flux} we show the transmitted
flux $T_f(E,\theta)$  as function of the energy and of the angle
$\theta$. The barrier between the leads and the center of the device
is represented by a delta function potential, therefore wider
barriers are represented by larger values of $\tilde g$. From
Fig.\ref{Fig_RTD_transmitted_flux} we can see that for larger values
of $\tilde g$ the transmission in the forward direction is
essentially zero except at some resonant energies, where the
transmission goes to one. As function of the angle we see that there
are some angles for which the transmission in also one. When the
length of central part of the device is increase (larger $d$) the
resonances become closer to each other and more resonances appear.

\begin{figure}
  \begin{center}
   \includegraphics[scale=.5]{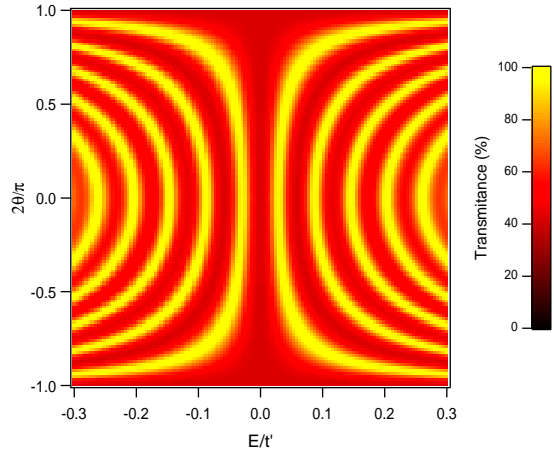}%
  \end{center}
\caption{(color on line) Intensity plot of $T_f(E,\theta)$ for a
device with $\tilde g=0.5$ and $d=200a$. There are clearly well
defined regions of large intensity  transmission.}
\label{Fig_intensity}
\end{figure}

In Figure \ref{Fig_intensity} we present an intensity plot of
$T_f(E,\theta)$ for a device  with $\tilde g=0.5$ and $d=200a$. In
this figure we can follow the evolution of the resonances in the $E$
versus $\theta$ plane. The six lines of larger transmission are
associated with the resonances we see in Fig.
\ref{Fig_RTD_transmitted_flux} for $\tilde g=2$ and $d=100a$.

\begin{figure}
  \begin{center}
   \includegraphics[scale=.5]{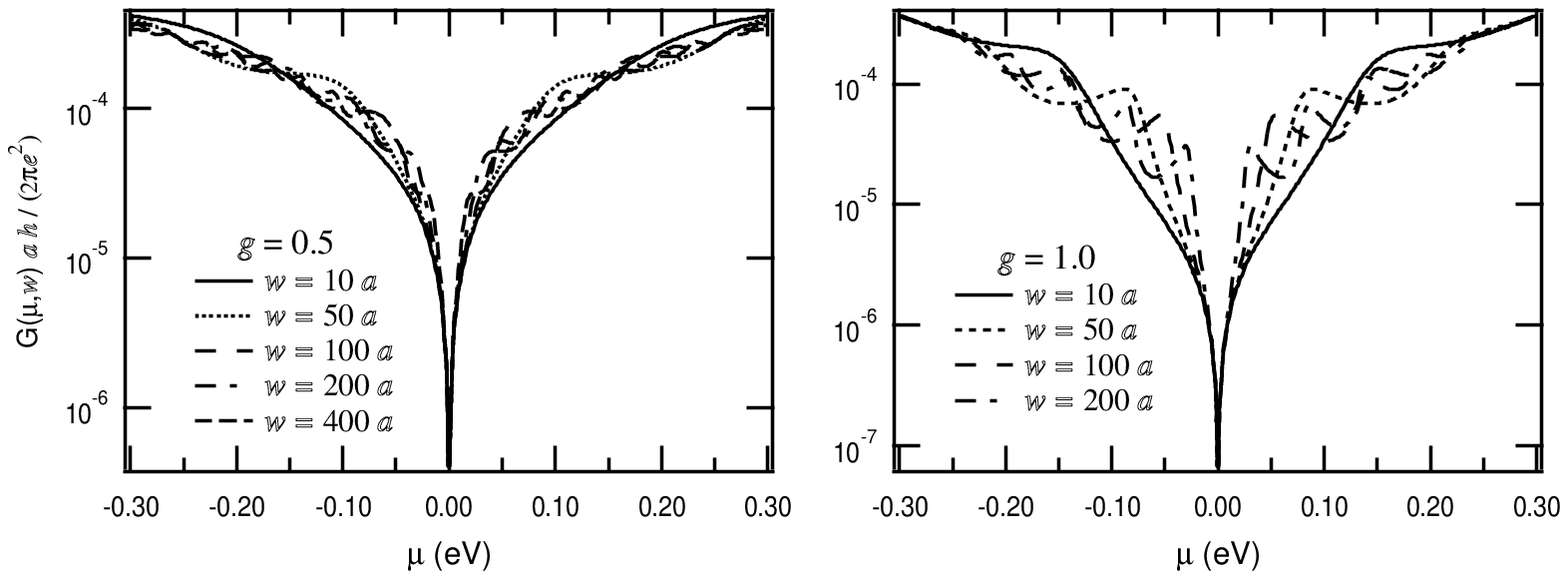}%
  \end{center}
\begin{center}
   \includegraphics[scale=.5]{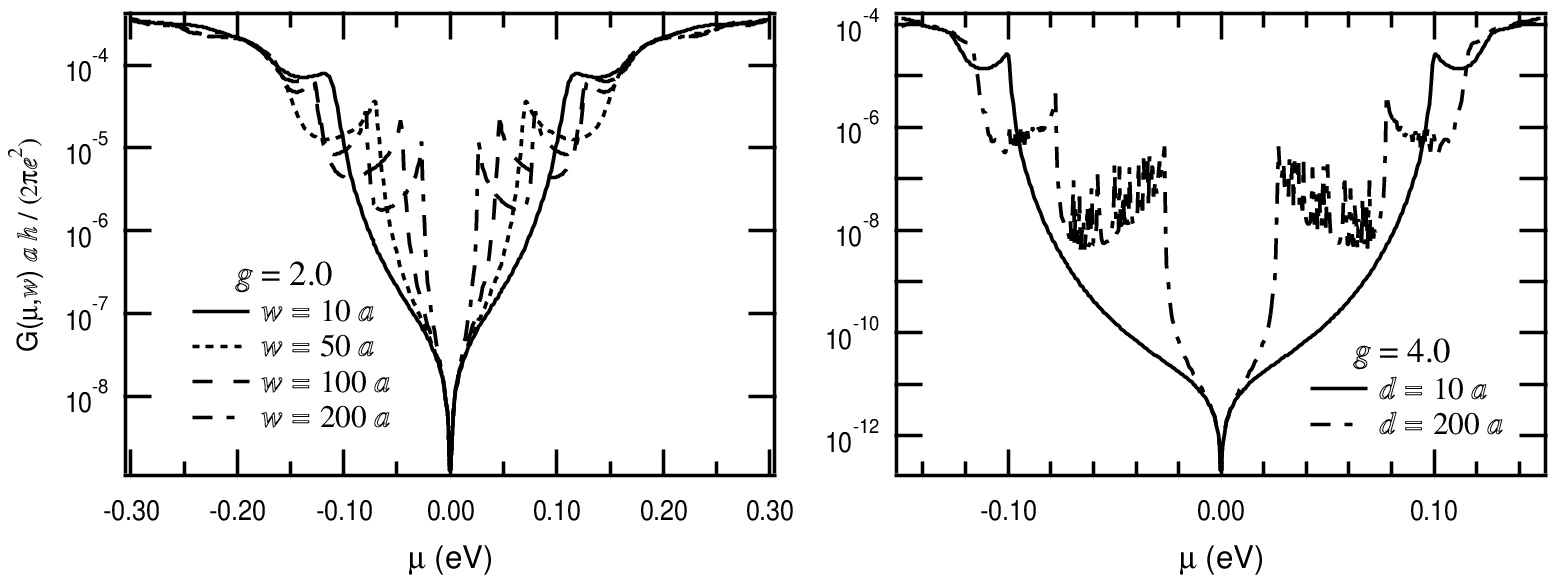}%
  \end{center}
\caption{Linear response conductance per unit of
cross-sectional length at zero temperature. The top panels show
$G(\mu,d)$ with $\tilde g$=0.5, 1 and the lower panels show the same
quantity but for $\tilde g$=2, 4.}
\label{Fig_G_linear_response_res_diode}
\end{figure}

As before, the linear response conductance per unit of
cross-sectional length is given, at zero temperature, by Eq.
(\ref{G}) and represented in Fig
\ref{Fig_G_linear_response_res_diode}. For small values of $\tilde
g$ the conductance shows smooth oscillations whereas for  larger
$\tilde g$ values strong resonances are observed. The number of
observed resonances depends on the length $d$.

%

\section{Final remarks.}

In this paper we discussed the tunneling properties of Dirac electrons in 
two-dimensions when they transverse regions of space where the spectrum
presents a finite energy gap. In the case we considered here the gap
is induced by depositing graphene on top of Boron Nitride, rendering, in this
way, the sub-lattices A and B non-equivalent. The consequence is the opening
of a gap in the energy spectrum, that we have parametrized by the parameter
$t'=0.1t$. We have shown that the existence of an energy gap prevents the
Klein paradox from taking place, a necessary condition for building 
nanoelectronic
devices made of graphene. We have also shown that basic devices like a resonant
tunneling diode can be made of graphene, by intercalating two regions where
the spectrum of graphene presents a gap. We have also shown that simple
analytical expressions can be derived for the tunneling through these types
of heterostructures. In addition we have showed that a limiting case of the
resonant tunneling diode  can be understood by using Dirac delta function
potentials.

Clearly that one is lead to think that a full description of the 
ballistic (no impurities) transport 
process in the system
should also include the effect of temperature and 
phonons.
 We note, however, that the
electron-phonon interaction has been shown to have a small effect
in the optical conductivity of graphene. \cite{Stauber} Which means that
phonons should not be very important in the description of the transport
process. 
Also the polaronic effect
leads to a renormalization of the velocity $v_F$ in the $k-region$ and
to the renormalization of the effective mass $mv^2_F=t'$ in the
$q-region$. But since these two parameters can be considered as effective 
ones there is no point in including the polaronic effect explicitly.
In concerns the temperature,
clearly it will be of no importance when the chemical potential
is above the gap. When the energy is in the gap there will certainly be
temperature activated transport adding on top of the tunneling current. For
small temperatures this will be a small effect.

 We believe our results of relevant for future nanoelectronics
applications of graphene.
%
\section*{Acknowledgments}
NMRP acknowledge the financial support from POCI
2010 via project PTDC/FIS/64404/2006. 

\appendix
\section{Details of section \ref{rtd} results.}
\label{app}
We now give the details that allows one to derive the result
\ref{tf}
Applying  boundary conditions derived in Sec. \ref{rtd} and a $T-$matrix
description\cite{Davies} for the scattering problem we obtain
\begin{equation}
\left(
\begin{array}{c}
A_{III}\\
B_{III}
\end{array}
\right) =T \left(
\begin{array}{c}
A_{I}\\
B_{I}
\end{array}
\right) = \left(
\begin{array}{cc}
T_{11}&T_{12}\\
T^\ast_{12}&T_{11}^\ast
\end{array}
\right) \left(
\begin{array}{c}
A_{I}\\
B_{I}
\end{array}
\right)
\end{equation}
with
\begin{equation}
T= V^{-1}_3M_2V^{-1}_2M_1\,. \label{T}
\end{equation}
The several matrices involved in Eq. (\ref{T}) are defined as
\begin{eqnarray}
M_1&=& \cosh\tilde g \left(
\begin{array}{cc}
 1 & 1\\
s_1e^{i\theta_1}&-s_1e^{-i\theta_1}
\end{array}
\right)\nonumber\\
&+&i\sinh\tilde g \left(
\begin{array}{cc}
s_1e^{i\theta_1}&-s_1e^{-i\theta_1}\\
 -1 & -1\\
\end{array}
\right)\,,
\end{eqnarray}

\begin{eqnarray}
V_2^{-1}&=& \frac 1 {2\cos\theta_2} \left(
\begin{array}{cc}
 e^{-i\theta_2}& s_2\\
 e^{i\theta_2}& -s_2
\end{array}
\right)\,,
\end{eqnarray}

\begin{eqnarray}
M_2&=& \cosh\tilde g \left(
\begin{array}{cc}
 e^{ik_2d} & e^{-ik_2d}\\
s_2e^{i(\theta_2+k_2d)}&-s_2e^{-i(\theta_2+k_2d)}
\end{array}
\right)\nonumber\\
&+&i\sinh\tilde g \left(
\begin{array}{cc}
s_2e^{i(\theta_2+k_2d)}&-s_2e^{-i(\theta_2+k_2d)}\\
 -e^{ik_2d} & -e^{-ik_2d}\\
\end{array}
\right) \,,
\end{eqnarray}
and
\begin{eqnarray}
V_3^{-1}&=& \frac 1 {2\cos\theta_3} \left(
\begin{array}{cc}
 e^{-i(\theta_3+k_3d})& s_3e^{-ik_3d}\\
 e^{i(\theta_3+k_3d)}& -s_3e^{ik_3d}
\end{array}
\right)\,.
\end{eqnarray}
The momenta $k_2$ and $k_3$ are given by
\begin{equation}
k_i=\frac {1}{v_F\hbar}\sqrt{(E-U_i)^2-(E-U_1)^2\sin^2\theta_1}\,,
\end{equation}
with $i=2,3$. The angles $\theta_i$ are defined as
\begin{equation}
\theta_i=\arctan\frac {k_y}{k_i}\,,
\end{equation}
with $k_y=\vert E-U_1\vert\sin\theta_1/(v_F\hbar)$. The $s_i$
functions are given by $s_i={\rm sign}(E-U_i)$, with $i=1,2,3$. The
potential energies $U_i$ represent some electrostatic potential
created in the corresponding region. Although  an analytical
expression for $T$ can be produced by carrying out the four matrix
multiplications, the resulting expression is too cumbersome to be
given here. In the special case that all $U_i=0$, the  matrix
elements of $T$ have a simple form given by

\begin{equation}
T_{11}=\cosh^2\tilde g +e^{-2idk}\sinh^2\tilde g\,,
\end{equation}
and
\begin{equation}
T_{12}=-sie^{-i\theta}(1+e^{-2ikd})\cosh\tilde g\sinh\tilde g\,,
\end{equation}
with $s={\rm sign}\,E$, $k=\vert E\vert \cos\theta/(v_F\hbar)$, and
$\theta$ the incident angle in the barrier.

%

\end{document}